\documentclass[twocolumn,runningheads]{svjour2}
\def\lesssim{\lower4pt\hbox{${\buildrel < \over \sim}$}}
\def\gtrsim{\lower4pt\hbox{${\buildrel > \over \sim}$}}
\smartqed  
\usepackage{graphicx}
\journalname{Astrophysics and Space Science}
\begin{document}

\title{Modeling the Emission Processes in Blazars
}


\author{Markus B\"ottcher
}


\institute{M. B\"ottcher \at
              Astrophysical Institute \\
              Department of Physics and Astronomy \\
              Ohio University \\
              Athens, OH 45701 \\
              USA \\
              Tel.: +1-740-593-1714\\
              Fax: +1-740-593-0433\\
              \email{mboett@helios.phy.ohiou.edu}           
}

\date{Received: date / Accepted: date}

\maketitle

\begin{abstract}
Blazars are the most violent steady/recurrent sources 
of high-energy gamma-ray emission in the known Universe. 
They are prominent emitters of electromagnetic radiation
throughout the entire electromagnetic spectrum. The 
observable radiation most likely originates in a relativistic
jet oriented at a small angle with respect to the line of
sight. This review starts out with a general overview of 
the phenomenology of blazars, including results from a 
recent multiwavelength observing campaign on 3C279. 
Subsequently, issues of modeling broadband spectra will
be discussed. Spectral information alone is not sufficient to 
distinguish between competing models and to constrain
essential parameters, in particular related to the primary
particle acceleration and radiation mechanisms in the jet. 
Short-term spectral variability information may help to
break such model degeneracies, which will require snap-shot
spectral information on intraday time scales, which may soon
be achievable for many blazars even in the gamma-ray regime
with the upcoming GLAST mission and current advances in
Atmospheric Cherenkov Telescope technology. In addition to
pure leptonic and hadronic models of gamma-ray emission from 
blazars, leptonic/hadronic hybrid models are reviewed, and 
the recently developed hadronic synchrotron mirror model 
for TeV $\gamma$-ray flares which are not accompanied by 
simultaneous X-ray flares (``orphan TeV flares'') is 
revisited. 
\end{abstract}

\section{Introduction}
\label{introduction}

Blazars (BL~Lac objects and $\gamma$-ray loud flat spectrum
radio quasars [FSRQs]) are the most extreme class of active
galaxies known. They have been observed at all wavelengths,
from radio through very-high energy (VHE) $\gamma$-rays.
46 blazars have been identified with high confidence as
sources of $> 100$~MeV emission detected by the {\it EGRET}
telescope on board the {\it Compton Gamma-Ray Observatory}
\cite{hartman99,mattox01}, and 
about one dozen blazars have now been detected at VHE 
$\gamma$-rays ($> 350$~GeV) by ground-based atmospheric 
\v Cerenkov telescopes (ACTs). Many of the {\it EGRET}-detected 
$\gamma$-ray blazars appear to emit the bulk of their bolometric 
luminosity at $\gamma$-ray energies. Blazars exhibit variability 
at all wavelengths on various time scales. Radio interferometry 
often reveals one-sided kpc-scale jets with apparent superluminal 
motion.

\begin{figure*}
\centering
\includegraphics[width=0.9\textwidth]{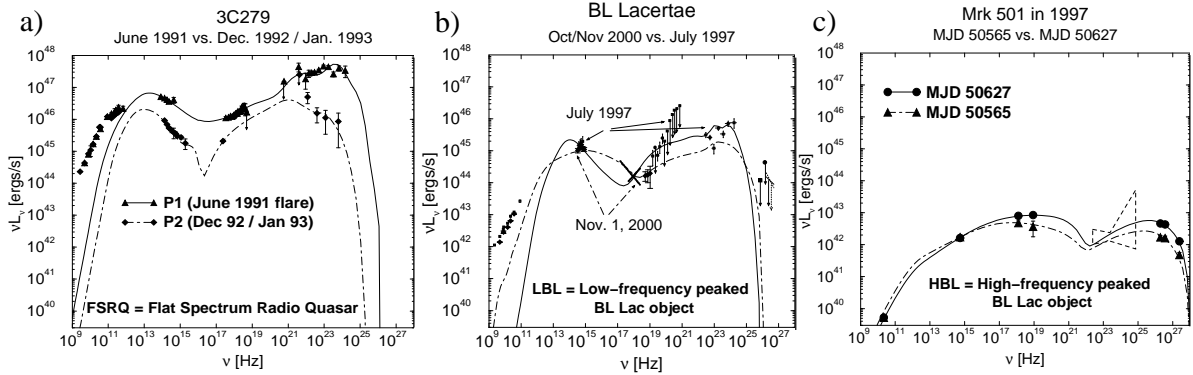}
\caption{\it SEDs of 3C~279 \cite{hartman01a}, BL~Lacertae
\cite{boettcher03}, and Mrk~501 \cite{petry00}. For each
object, two simultaneous broadband spectra at two different epochs are
shown. The curves show model fits, using a leptonic jet model.}
\label{bbspectra}
\end{figure*}

\subsection{Spectral classification of blazars}
\label{classification}

The broadband continuum spectra of blazars are dominated by non-thermal
emission and consist of two distinct, broad components: A low-energy
component from radio through UV or X-rays, and a high-energy component
from X-rays to $\gamma$-rays. A sequence of blazar sub-classes, from 
FSRQs to low-frequency peaked BL Lac objects (LBLs) to high-frequency 
peaked BL Lacs (HBLs) can be defined through the peak frequencies 
and relative $\nu F_{\nu}$ peak fluxes, which also seem to be 
correlated with the bolometric luminosity \cite{fossati98}. 
The sequence FSRQ $\to$ LBL $\to$ HBL is characterized by increasing
$\nu F_{\nu}$ peak frequencies, a decreasing dominance of the $\gamma$-ray
flux over the low-frequency emission, and a decreasing bolometric 
luminosity (see Fig. 1). LBLs are intermediate between the FSRQs and 
the HBLs. The peak of their low-frequency component is located at IR 
or optical wavelengths, their high-frequency component peaks at several 
GeV, and the $\gamma$-ray output is of the order of or slightly higher 
than the level of the low-frequency emission. However, the existence 
and physical significance of this blazar sequence has recently been 
questioned and the apparent sequence attributed to selection effects 
due to the use of flux-limited samples (for a recent review see, e.g., 
\cite{padovani06}).

\begin{figure*}
\centering
\includegraphics[width=0.9\textwidth]{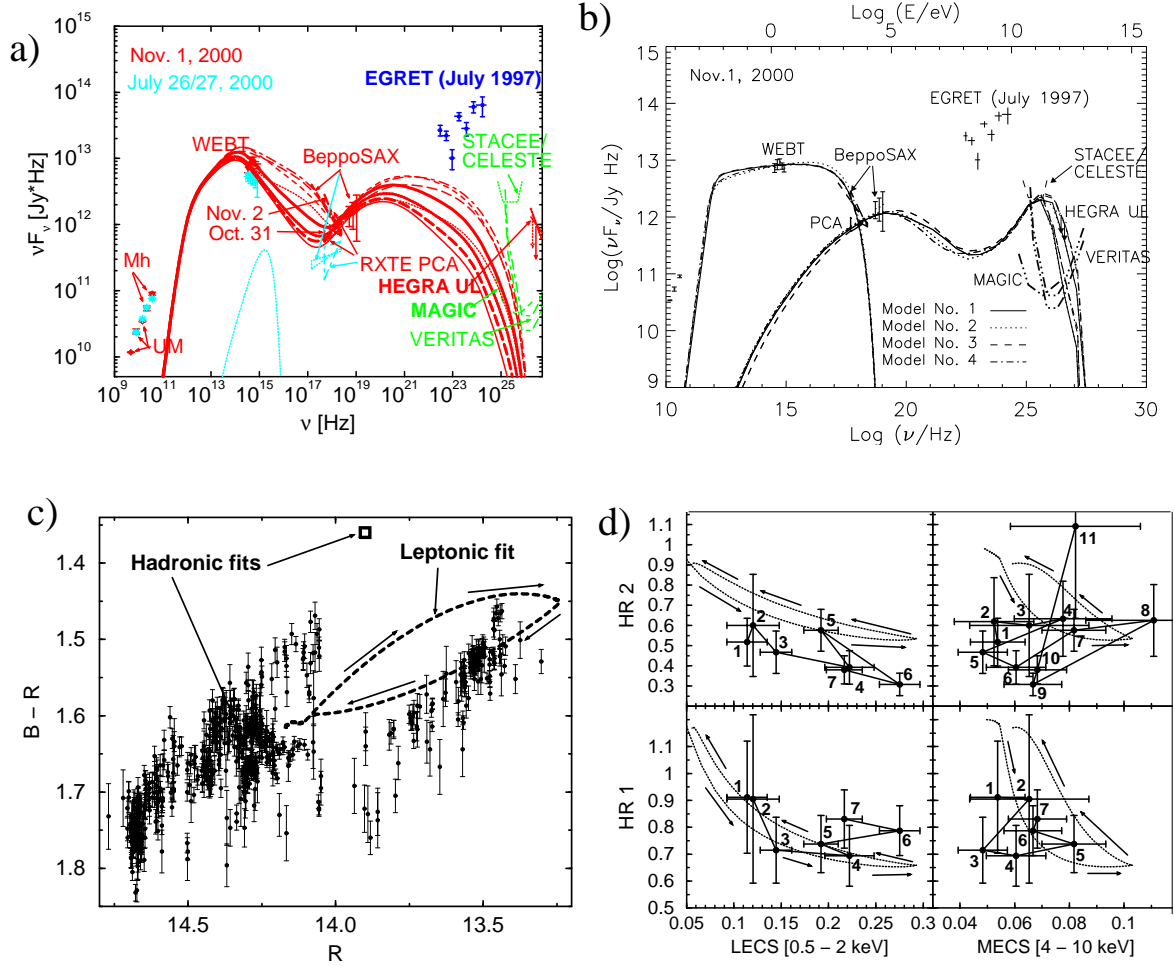}
\caption{Spectral variability fitting of BL~Lacertae in 2000
\cite{br04}: (a) Time-dependent leptonic fits to the 
Nov. 1 high state; (b) Various hadronic fits, differing mainly 
in their co-moving magnetic-field and synchrotron photon energy 
density. (c) Comparison of the simulation results corresponding 
to fits from panel (a) and (b) to the optical color-magnitude 
correlation; (d) comparison of the leptonic fit results to the 
X-ray hardness-intensity diagrams during a short flare observed 
by {\it BeppoSAX} on Nov. 1. Based on count rates in the three 
{\em BeppoSAX} NFI energy channels LECS [0.5 -- 2 keV], MECS 
[2 -- 4 keV], and MECS [4 -- 10 keV], and X-ray hardness ratios: 
HR1 = MECS[2-4]/LECS[0.5-2], HR2 = MECS[4-10]/MECS[2-4]. }
\label{bllac_var}
\end{figure*}

\subsection{Spectral variability of blazars}
\label{variability}

Fig. \ref{bbspectra} already illustrates that in particular the 
high-energy emission from blazars can easily vary by more than 
an order of magnitude between different {\it EGRET} observing 
epochs \cite{vmontigny95,mukherjee97,mukherjee99}. However, 
high-energy variability has been observed on much shorter 
time scales, in some cases less than an hour \cite{gaidos96}.

BL~Lac objects occasionally exhibit X-ray variability patterns which can 
be characterized as spectral hysteresis in hardness-intensity diagrams 
(e.g., \cite{takahashi96,kataoka00,fossati00,zhang02}). 
This has been interpreted as the synchrotron signature of gradual
injection and/or acceleration of ultrarelativistic electrons in 
the emitting region, and subsequent radiative cooling (e.g., 
\cite{kirk98,gm98,kataoka00,kusunose00,li00,bc02}).
While spectral hysteresis has so far only been clearly identified
in HBLs, it should also occur in the soft X-ray emission of LBLs if
their synchrotron component extends into the soft X-ray regime.
However, LBLs are generally fainter at X-ray energies than HBLs,
making the extraction of time-resolved spectral information
observationally very challenging. Fig. \ref{bllac_var}d shows
the results of a {\em BeppoSAX} observation of BL~Lacertae in 2000
\cite{ravasio03,boettcher03}. Rapid flux and spectral variability 
of blazars is also commonly observed in the optical regime, often 
characterized by a spectral hardening during flares (see, e.g., 
Fig. \ref{bllac_var}c, or \cite{lainela99,villata02}).

\section{\label{3C279}Preliminary results from 3C279 2006}

The flat spectrum radio quasar 3C279 ($z = 0.538$) is one of the 
best observed flat spectrum radio quasars, not at last because of 
its prominent $\gamma$-ray flare shortly after the launch of the 
{\it Compton Gamma-Ray Observatory} (CGRO) in 1991. It has been 
persistently detected by the Energetic Gamma-Ray Experiment Telescope 
(EGRET) on board CGRO each time it was observed, even in its very 
low quiescent states, e.g., in the winter of 1992 -- 1993, and 
is known to vary in $\gamma$-ray flux by roughly two orders of 
magnitude \cite{maraschi94,wehrle98}. It has been monitored
intensively at radio, optical, and more recently also X-ray
frequencies, and has been the subject of intensive multiwavelength
campaigns (e.g., \cite{wehrle98}). Its broadband SEDs at several 
epochs are rather well determined, but a complete compilation and
modeling (using a leptonic model, see \S \ref{leptonic}) of
all available SEDs simultaneous with the 11 EGRET observing epochs
yielded somewhat inconclusive results \cite{hartman01a}. Furthermore, 
in spite of the intensive past observational efforts, the broadband 
spectral variability of 3C~279 is still rather poorly understood 
(see, e.g., \cite{bednarek98,hartman01b,sikora01,moderski03}). 

\begin{figure}
\centering
\includegraphics[width=0.45\textwidth]{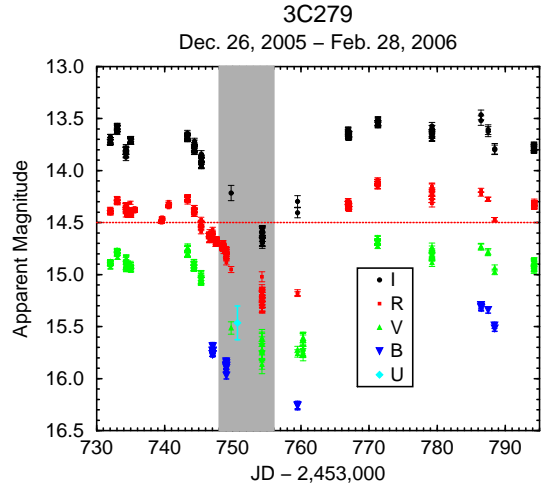}
\caption{Preliminary optical (BVRI) light curves of 3C~279 over the
entire core period of the 2006 multiwavelength campaign. The shaded
area indicates the period of the INTEGRAL X-ray and soft $\gamma$-ray
observations. The red horizontal line indicates the trigger criterion
of R = 14.5.}
\label{3C279_optical}
\end{figure}

For the reasons stated above, we \cite{collmar06} proposed an
intensive multiwavelength campaign in an optical high state of
the 3C~279, in order to investigate its correlated radio -- IR
-- optical -- X-ray -- soft $\gamma$-ray variability. The campaign
was triggered on Jan. 5, 2006, when the source exceeded an R-band 
flux corresponding to R = 14.5. It involved intensive radio, near-IR
(JHK), and optical monitoring by the WEBT collaboration through 
March of 2006, focusing on a core period of Jan. and Feb. 2006. 
X-ray and soft $\gamma$ observations were carried out by all 
instruments on board the {\it International Gamma-Ray Astrophysics
Laboratory} (INTEGRAL) during the period of Jan. 13 -- 20, 2006.
Additional, simultaneous X-ray coverage was obtained by {\it Chandra}
and {\it Swift} XRT. These observations were supplemented by extended
X-ray monitoring with the {\it Rossi X-Ray Timing Explorer} (RXTE)
and VLBA monitoring at 43~GHz. The analysis of the data collected
during this campaign is currently in progress. Here, first, 
preliminary results from this campaign are presented. Final
analysis results of the WEBT (radio -- IR -- optical) campaign will
be published in \cite{boettcher06}, while a comprehensive report on
the result of the entire multiwavelength campaign will appear in
\cite{collmar06}.

\begin{figure}
\centering
\includegraphics[width=0.45\textwidth]{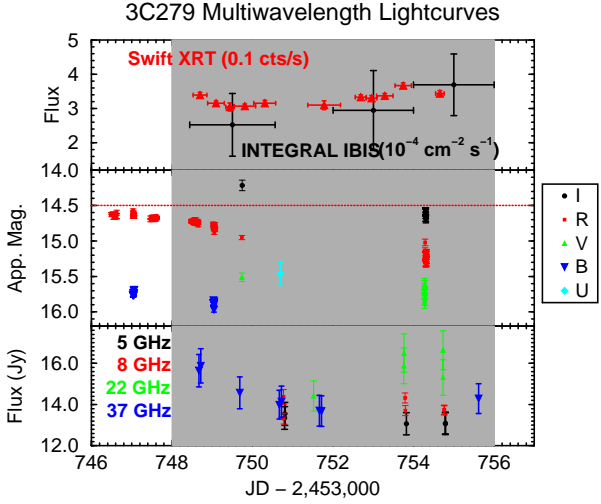}
\caption{Preliminary X-ray and soft $\gamma$-ray (top panel), 
optical (middle panel), and radio (bottom panel) light curves 
of 3C~279 during the time of the INTEGRAL observations on Jan. 13
-- 20, 2006.}
\label{3C279_mw_lc}
\end{figure}

Fig. \ref{3C279_optical} shows the preliminary optical (BVRI) light
curves, including about 3/4 of all collected data from the campaign.
The figure illustrates that the source showed substantial, closely
correlated variability in all wave bands throughout the entire
core campaign. Between Jan. 8 and 15 (i.e., including the time of
the INTEGRAL observations), the optical flux was persistently fading
in all optical bands so that the R band flux was actually below the 
intended trigger threshold. However, this may not necessarily be bad
news since this might allow us to study the correlated radio through
X-ray flux decay time scales during a period of clean, steady decay after 
a major optical outburst. We might thus be able to probe energy-dependent
electron cooling time scales throughout the entire synchrotron
component and the low-energy part of the high-energy component
(generally attributed to synchrotron-self-Compton emission in
leptonic jet models, e.g., \cite{hartman01a}) of the SED of 3C279.
Fig. \ref{3C279_mw_lc} illustrates that the general trend of flux
decline throughout most of the time window of the INTEGRAL observations 
was also shared by the high-frequency (37~GHz) radio flux, while the
X-ray flux detected by Swift indicates a slowly rising trend after 
Jan. 14. Possible interpretations of this behavior will be discussed
after the brief introduction to leptonic blazar jet models in \S 
\ref{leptonic}.

\begin{figure}
\centering
\includegraphics[width=0.45\textwidth]{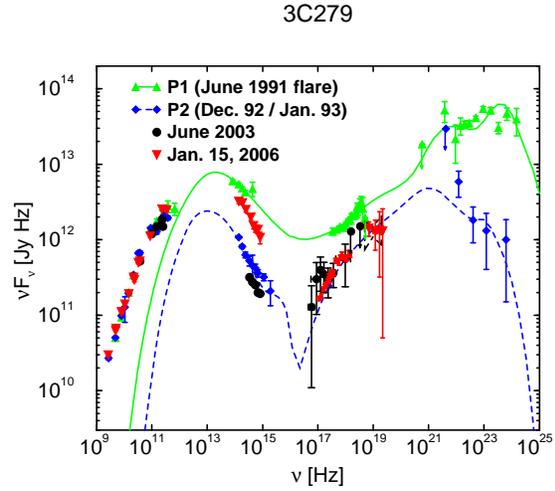}
\caption{Simultaneous spectral energy distributions of 3C~279
during 4 observing epochs, including a snapshot SED on Jan. 15,
2006 (red triangles pointing down).}
\label{3C279_SED}
\end{figure}

In Fig. \ref{3C279_SED}, we compare a snapshot SED of 3C~279
of Jan. 15, 2006 to several other simultaneous SEDs, including
the exceptional June 1991 flare and the very low, quiescent
state of Dec. 1992 -- Jan. 1993, as well as the time of the
previous INTEGRAL observation in June 2003 \cite{collmar04}. 
It reveals a surprising result: Even though the optical flux
is well (factor 2 -- 3) above the typical low-state values,
the X-ray and soft $\gamma$-ray fluxes and spectra are in 
the range of the lowest X-ray fluxes measured from the source, 
including the 1992/93 and 2003 quiescent states. Before we
procede with a possible physical interpretation of the results
from this campaign, let us briefly review the salient features
of blazar jet models. 

\section{\label{models}Blazar models}

The high inferred bolometric luminosities, rapid variability,
and apparent superluminal motions provide compelling evidence
that the nonthermal continuum emission of blazars is produced
in $\lesssim$~1 light day sized emission regions, propagating
relativistically along a jet directed at a small angle with 
respect to our line of sight. It is generally agreed that 
the low-frequency component of blazar SEDs might be synchrotron 
radiation from nonthermal, ultrarelativistic electrons.
Several electron injection/acceleration scenarios have been
proposed, e.g. impulsive injection near the base of the jet
(e.g., \cite{ds93,dermer97}; such a scenario might also apply
to originally Poynting-flux dominated jets, see 
\cite{sikora05}), isolated shocks propagating along 
the jet (e.g., \cite{mg85,kirk98,sikora01,sokolov04}), 
internal shocks from the collisions of multiple shells 
of material in the jet \cite{spada01}, stochastic particle 
acceleration in shear boundary layers of relativistic jets 
(e.g., \cite{ob02,rd04}), magnetic reconnection in Poynting-flux
dominated jets (e.g., \cite{sikora05}), or hadronically 
initiated pair avalanches \cite{km99}. 

Significant progress has recently been made in our understanding 
of particle acceleration at relativistic shocks (e.g., 
\cite{achterberg01,ob02,no04,vv05}) and the conversion of 
relativistic bulk kinetic energy into relativistic particles 
and ultimately into radiation \cite{ps00,schlickeiser02,vainio04,mk06}.
While particle acceleration at relativistic parallel shocks might 
produce electron injection spectra of $\dot N_e (\gamma) \propto 
\gamma^{-q}$ with $2.2 \lesssim q \lesssim 2.3$
(e.g., \cite{achterberg01}), oblique shocks tend to produce much 
softer injection spectral indices (e.g., \cite{ob02}). In contrast, 
the effect of stochastic acceleration in resonance with plasma
wave turbulence behind the shock front might harden the injection 
index significantly, possibly even beyond $q \sim 1$ \cite{vv05}. 
In the framework of an inhomogeneous jet with a fast inner spine 
and a slower, outer cocoon, particle acceleration at shear boundary 
layers may become the dominant acceleration mechanism 
\cite{ostrowski00,so02,rd04}. This may lead to the 
development of relativistic particle spectra with indices 
of $q < 2$ and a high-energy bump around the energy where 
the particle acceleration rate equals the energy loss rate. 

However, the lack of knowledge of the primary jet launching mechanism 
(poynting-flux dominated vs. mag\-ne\-to-hy\-dro\-dyn\-am\-ic, see, e.g., 
\cite{sikora05}) and the primary pair loading of the jet are currently severe 
problems in tying the properties of the particle acceleration mechanism 
to more fundamental physical properties of the accretion-powered disk-jet 
system. There is mounting evidence that --- if the high-energy emission 
of blazars is leptonically dominated --- jets of blazars might be 
energetically and dynamically dominated by their proton content, though 
pairs may still greatly outnumber protons \cite{sm00,gc01,kt04,sikora05}

While the electron-synchrotron origin of the low-fre\-quen\-cy emission 
is well established, there are two fundamentally different approaches
concerning the high-energy emission. If protons are not accelerated
to sufficiently high energies to reach the threshold for $p\gamma$ pion
production on synchrotron and/or external photons, the high-energy radiation
will be dominated by emission from ultrarelativistic electrons and/or 
pairs (leptonic models). In the opposite case, the high-energy emission 
will be dominated by cascades initiated by $p\gamma$ pair and pion 
production as well as proton, $\pi^{\pm}$, and $\mu^{\pm}$ synchrotron 
radiation (hadronic models). 

\subsection{\label{leptonic}Leptonic blazar models}

If protons are not accelerated to sufficiently high energies to reach 
the threshold for $p\gamma$ pion production on synchrotron and/or 
external photons, the high-energy emission will be dominated by 
ultrarelativistic electrons and/or pairs. In this case, high-energy 
emission can be produced via Compton scattering off the same 
ultrarelativistic electrons producing the synchrotron 
emission at lower frequencies. Possible target photon fields are 
the synchrotron photons produced within the jet (the SSC process: 
\cite{mg85,maraschi92,bm96}) or external photons (the EC process). 
Possible sources of external seed photons are accretion-disk photons
entering the jet directly \cite{ds93} or after being reprocessed in
the circumnuclear material (e.g., the broad line regions of quasars)
\cite{sikora94,dermer97}, jet synchrotron emission reflected off 
clouds in the circumnuclear material \cite{gm96}, infrared emission
from a dust torus around the central engine \cite{blazejowski00,arbeiter02}.
In addition, $\gamma\gamma$ absorption, pair production, and synchrotron 
self absorption must be taken into account in a self-consistent leptonic 
blazar model. As the emission region is propagating relativistically 
along the jet, continuous particle injection and/or acceleration and 
subsequent radiative and adiabatic cooling, particle escape, and 
possibly also the deceleration of the jet, in particular in HBLs 
\cite{gk03,ghisellini05}, have to be considered. 

As one may expect from the basic description in the previous paragraph, 
leptonic models (especially when considering a substantial contribution
from external radiation fields as targets for Compton upscattering to
$\gamma$-rays) require the specification of a rather large number of 
parameters. Several of these parameters can be estimated to a reasonable 
degree of accuracy from observables (see, e.g., \cite{boettcher03} for 
a discussion of such parameter estimates in the case of BL~Lacertae
during the multiwavelength campaign of 2000). In particular, there
are several ways to obtain an estimate on the magnetic field in the
emission region, which is an essential ingredient to also get a handle
on the efficiency of Fermi and stochastic acceleration of particles 
in the jet. 

One way to estimate the co-moving magnetic field can be 
found by assuming that the dominant portion of the time-averaged 
synchrotron spectrum is emitted by a quasi-equilibrium power-law 
spectrum of electrons with $N_e (\gamma) = n_0 \, V_B \, \gamma^{-p}$
for $\gamma_1 \le \gamma \le \gamma_2$; here, $V_B$ is the co-moving
blob volume. The normalization constant $n_0$ is related to the
magnetic field through an equipartition parameter $e_B \equiv u_B
/ u_e$ (in the co-moving frame). Note that this equipartition parameter
only refers to the energy density of the electrons, not accounting for
a (possibly greatly dominant) energy content of a hadronic matter 
component in the jet. Under these assumptions, the $\nu F_{\nu}$
peak synchrotron flux $f_{\epsilon}^{\rm sy}$ at the dimensionless 
synchrotron peak energy $\epsilon_{\rm sy} = h \nu_{\rm sy} / 
(m_e c^2)$ is approximately given by

\begin{equation}
f_{\epsilon}^{\rm sy} = (D \, B)^{7/2} \, {\pi \, c \, \sigma_{\rm T} 
\over 288 \, d_L^2} \, \left( [1 + z] \, \epsilon_{\rm sy} \, B_{\rm cr} 
\right)^{1/2} \, { p - 2 \over e_B \, m_e c^2}
\label{f_sy}
\end{equation}
where $D = 10 \, D_1 = (\Gamma [1 - \beta_{\Gamma} \cos\theta_{\rm obs}])^{-1}$ 
is the Doppler boosting factor, $d_L = 10^{27} \, d_{27}$~cm is the the
luminosity distance of the source, and $B_{\rm cr} = 4.414 \times 10^{13}$~G. 
The electron spectrum normalization used to derive Eq. \ref{f_sy} is based 
on the presence of a power-law shape with a photon energy index $\alpha > 1$ 
(with $F_{\nu} \propto \nu^{-\alpha}$) of the synchrotron spectrum beyond 
the synchrotron peak. If the SED reveals such a spectral shape, the 
underlying electron spectrum always has an index of $p \ge 3$. Eq. \ref{f_sy} 
then yields a magnetic-field estimate of

\begin{equation}
B_{e_B} = 9 \, D_1^{-1} \left( {d_{27}^4 \, f_{-10}^2 \, e_B^2 \over
[1 + z]^4 \, \epsilon_{\rm sy, -6} \, R_{15}^6 \, [p - 2]} \right)^{1/7}
\; {\rm G},
\label{B_equipartition}
\end{equation}
where $f_{-10} = f_{\epsilon}^{\rm sy}$/(10$^{-10}$~ergs~cm$^{-2}$~s$^{-1}$),
$\epsilon_{\rm sy, -6} = \epsilon_{\rm sy}/10^{-6}$, and $R_B = 10^{15} \,
R_{15}$~cm is the transverse radius of the emission region. The bulk 
Lorentz factor $\Gamma$ and Doppler factor $D$ can usually be constrained 
from superluminal motion measurements and/or from constraints on the 
compactness of the emission region, which is an approximate measure of 
the optical depth to $\gamma\gamma$ absorption. An estimate of the size 
of the emission region can be inferred from the minimum variability time
scale, $t_{\rm var} = t_{\rm var, h}$~hr, as $R_B \sim 10^{15} \, D_1 \, 
t_{\rm var, h}$~cm. 

We can apply this estimate to the results of our 2006 campaign on 
3C~279 described in \S \ref{3C279}. For $z = 0.538$, $H_0 = 
70$~km~s$^{-1}$~Mpc$^{-1}$, $\Omega_m = 0.3$, and $\Omega_{\Lambda}
= 0.7$, the luminosity distance is $d_L = 9.3 \times 10^{27}$~cm.
From previous work on 3C~279 (see, e.g., \cite{hartman01a} for a 
summary) one finds typical values of $D \sim 10$, and $R_B \sim 6 
\times 10^{16}$~cm. Unfortunately, the synchrotron peak seems to
lie in the mm -- far-IR regime of the spectrum, which was not
covered during the campaign. Thus, its position is not very well 
constrained. Visual inspection of the SED suggest values of 
$f_{-10} \approx 0.7$ and $\nu_{\rm sy} \approx 8 \times 10^{13}$~Hz, 
corresponding to $\epsilon_{\rm sy, -6} \approx 0.65$. The IR -- 
optical (UBVRIJHK) spectral index is $\alpha = 1.75$, which 
corresponds to $p = 4.5$. These values lead to a magnetic-field
estimate of 

\begin{equation}
B_{e_B} \sim 0.6 \, D_1^{-1} \, e_B^{2/7} \; {\rm G}.
\label{B_eB}
\end{equation}
which appears to be relatively weak compared to typical values
of $\sim$~a few G inferred from modeling efforts on other FSRQs.
The steep spectral index suggests that electrons emitting IR and 
optical radiation might already be in the strong cooling regime, 
where the radiative cooling time is shorter than the dynamical
time scale. Even in that case, one would infer that electrons
are injected through the primary particle acceleration mechanism
with an injection index of $q \sim 3.5$, which would imply rather
inefficient acceleration, and may point towards Fermi acceleration
at oblique shocks as the primary particle acceleration mechanism
(see the discussion in \S \ref{models}).

Although several model parameters of blazar jet models can 
be reasonably well constrained from the broadband spectral 
properties of blazars, spectral fitting alone is generally 
insufficient to constrain all relevant model parameters 
(see, e.g., the analysis in \cite{boettcher02} for the case 
of the LBL W~Comae). Thus, it is now widely agreed that
spectral and variability properties of blazars have to be taken
into account simultaneously in order to extract as much physical
information as possible from simultaneous multiwavelength observing
campaigns. 

Significant progress has been made in the past few years to combine 
spectral and variability modeling of blazars using leptonic models. 
In particular, the spectral variability of HBLs has been modelled 
in great detail by many authors, using pure SSC models (e.g., 
\cite{gm98,kataoka00,krawczynski02,sokolov04}. Time-dependent 
blazar modeling including external soft photon sources is also 
advancing rapidly (e.g., \cite{sikora01,bc02,kusunose03,sm05}), 
providing tools for the interpretation of simultaneous spectral
and variability data from LBLs and FSRQs. An instructive example 
of combined fitting of SEDs and rapid spectral variability of
BL~Lacertae can be found in \cite{br04} (see also Fig. \ref{bllac_var}).

In this context, one should mention an alternative way of estimating 
the magnetic field in blazar jets. This is based on a possible time 
delay between light curves at two different frequencies at which the
emission is dominated by synchrotron emission. Assuming that such a 
delay is caused by synchrotron cooling of high-energy electrons with
characteristic observed synchrotron photon energy $E_{\rm sy, 0} =
E_0$~keV to lower energies with corresponding synchrotron energy
$E_{\rm sy, 1} = E_1$~keV, the magnetic field can be estimated as:

\begin{eqnarray}
\label{B_delay}
B_{\rm delay} = & 0.4 \, D_1^{-1/3} \, (1 + k)^{-2/3} \, 
(\Delta t_h^{\rm obs})^{-2/3} \cr\cr
& (E_1^{-1/2} - E_0^{-1/2})^{2/3} \; {\rm G}.
\end{eqnarray}
where $k = u_{\rm ph} / u_{\rm B}$ is the ratio of energy densities
in the photon field in the frame co-moving with the emission
region and the magnetic field, and $\Delta t_h^{\rm obs}$ is 
the observed time delay in hours. 

Our preliminary analysis of the results from the 2006 campaign on
3C~279 did not yield any evidence for time lags between any of the
near-IR and optical bands. Furthermore, the SED of Jan. 15 (see
Fig. \ref{3C279_SED}) clearly shows that the X-ray emission is
dominated by the low-energy end of the high-energy spectral component.
Therefore, possible X-ray --- optical delays can not be used for 
the estimate in Eq. \ref{B_delay}. 

The somewhat surprising apparent trend of a slow rise in the X-ray
and soft $\gamma$-ray fluxes during the continuing optical fading
trend illustrated in Fig. \ref{3C279_mw_lc} may be interpreted in
the following way: In leptonic jet models, the low-energy end
of the high-frequency bump of the SEDs of blazars (which covers the
X-ray band in the case of 3C~279) is generally attributed to synchrotron
self-Compton emission from rather low-energy electrons. The differential 
number density of electrons at moderately relativistic energies will be
gradually built up through the relatively long electron cooling time
scale of electrons at those energies. In addition, the self-generated 
synchrotron photon field, serving as a target for the SSC process will 
also be gradually built up throughout at least the dynamical time scale.
Consequently, substantial time delays of the X-ray emission behind
optical flaring activity may result (see, e.g., \cite{sikora01} for
a detailed discussion). This could mean that the slow rise of the
X-ray flux is, in fact, the delayed response to the optical flare
observed around Jan. 8, 2006. A more quantitative analysis of this
interpretation will be presented in \cite{collmar06}.

\subsection{\label{hadronic}Hadronic blazar models}

If a significant fraction of the kinetic power in the jet is converted
into the acceleration of relativistic protons and those protons reach the
threshold for $p\gamma$ pion production, synchrotron-supported pair 
cascades will develop \cite{mb92,mannheim93}. The acceleration of protons 
to the necessary ultrarelativistic energies requires high magnetic fields 
of at least several tens of Gauss. In the presence of such high magnetic
fields, the synchrotron radiation of the primary protons 
\cite{aharonian00,mp00} and of secondary muons and mesons 
\cite{rm98,mp00,mp01,muecke03} must be taken into account 
in order to construct a self-consistent synchrotron-proton 
blazar (SPB) model. Electromagnetic cascades can be
initiated by photons from $\pi^0$-decay (``$\pi^0$ cascade''), electrons
from the $\pi^\pm\to \mu^\pm\to e^\pm$ decay (``$\pi^\pm$ cascade''),
$p$-synchrotron photons (``$p$-synchrotron cascade''), and $\mu$-, $\pi$-
and $K$-synchrotron photons (``$\mu^\pm$-synchrotron cascade'').

It has been shown in \cite{mp01,muecke03} that the 
``$\pi^0$ cascades'' and ``$\pi^\pm$ cascades'' generate featureless
$\gamma$-ray spectra, in contrast to ``$p$-synchrotron cascades'' and
``$\mu^\pm$-syn\-chro\-tron cascades'' that produce a double-bumped 
$\gamma$-ray spectrum. In general, direct proton and $\mu^{\pm}$ 
synchrotron radiation is mainly responsible for the high energy 
bump in blazars, whereas the low energy bump is dominanted by 
synchrotron radiation from the primary $e^-$, with a contribution 
from secondary electrons. Fig. \ref{bllac_var}b shows fits to the 
SED of BL~Lacertae in 2000, using the hadronic SPB model \cite{mp01}.

\subsection{\label{unification}Blazar unification}

Leptonic models have been used successfully to reproduce simultaneous
SEDs of several blazars. Spectral modeling results are now converging
towards a rather consistent picture \cite{ghisellini98,kubo98}.
The sequence HBL $\to$ LBL $\to$ FSRQ appears to be related
to an increasing external-Compton contribution to the $\gamma$-ray
spectrum. While most FSRQs are successfully modelled with EC models
(e.g., \cite{dermer97,sambruna97,mukherjee99,hartman01a}), the 
SEDs of HBLs are consistent with pure SSC models (e.g., 
\cite{mk97,pian98,petry00,krawczynski02}). LBLs (e.g., BL~Lacertae, 
W~Comae) often seem to require an EC component to explain their {\it EGRET} 
spectra \cite{sambruna99,madejski99,bb00,boettcher02}. One generally 
finds that HBLs require higher average electron energies and lower 
magnetic fields than LBLs and FSRQs. In the framework of a unified 
leptonic model, this basic parameter sequence may be related to an 
increasing importance of EC cooling along the sequence HBL $\to$ LBL 
$\to$ FSRQ \cite{ghisellini98}. It has been suggested 
that the decreasing importance of external radiation fields along 
the sequence FSRQ $\to$ LBL $\to$ HBL may be an evolutionary effect 
related to the gradual depletion of a limited reservoir of circumnuclear 
material \cite{dc01,cd02,bd02}.

Hadronic blazar models also offer a physical interpretation for 
the spectral sequence of BL Lac subclasses \cite{muecke03}. The
spectra of HBLs are well reproduced by p-synchrotron dominated SPB 
models where the intrinsic primary synchrotron photon energy density
is small, consistent with the low bolometric luminosity of those objects. As
the synchrotron photon energy density increases towards LBL-like synchrotron
properties, protons suffer increasingly strong $p\gamma$ pion production
losses, and the contributions from the $\pi^{\pm}$ and $\mu^{\pm}$ synchrotron
cascades become increasingly dominant at higher energies. This results in
a decreasing $\nu F_{\nu}$ peak frequency of the $\gamma$-ray component.
The effect of external photon sources might further enhance the
$\pi^{\pm}$-synchrotron and $\mu^{\pm}$-synchrotron cascade contributions,
reproducing the transition to quasar-like properties.

It should be pointed out here that the blazar sequence, if real, can
be explained in the framework of both leptonic and hadronic models,
but it is {\it not} a prediction of either one of the classes of
models. Consequently, even if future observations reveal evidence
for a rather uniform distribution of peak frequencies and relative
luminosities between the two main spectral components of blazars,
both model variants remain generally viable.

\subsection{\label{hybrid}Hybrid blazar models}

The leptonic and hadronic models discussed above are certainly only
to be regarded as extreme idealizations of a blazar jet. Realistically,
both types of processes might play a role to some extent and should 
thus be considered to a comparable level of sophistication. The 
recent observation of isolated TeV flares without simultaneous 
X-ray flares (a phenomenon sometimes referred to as ``orphan TeV 
flares'') in 1ES~1959+650 \cite{krawczynski04} and Mrk~421 
\cite{blazejowski05} may, in fact, provide rather strong support 
for the importance of hadronic processes in objects of which other 
spectral and variability features are generally well reproduced 
by leptonic jet models, since standard leptonic SSC models predict 
a close temporal flux correlation between the synchrotron and 
Compton components. In 1ES~1959+650, the ``orphan'' TeV flare 
was preceded by an ordinary, correlated X-ray and TeV-flare, 
which can be generally well understood in the context of leptonic 
SSC models. This finding strongly suggests the need for models 
that explain flares dominated by leptonic interactions as well as 
flares where non-leptonic components might play an important role 
within the same system. 

Hadronic processes in the context of models with leptonically dominated 
blazar emission have been considered by several authors, e.g.:

\begin{itemize}

\item{A "supercritical pile" model was suggested in \cite{km99}. In 
this model a runaway pair production avalanche is initiated by mildly 
relativistic protons interacting with reflected synchrotron photons 
via $p\gamma$ pair production, as the primary pair injection mechanism 
in blazar jets. Spectral characteristics resulting from this model
as applied to gamma-ray bursts have been considered in \cite{mk06}.}

\item{The conversion of ultrarelativistic protons into neutrons 
via $p\gamma$ pion production on external soft photons was suggested 
in \cite{ad03} as a mechanism to overcome synchrotron losses of protons 
near the base of blazar jets and, thus, to allow blazar jets to remain 
collimated out to kpc scales.}

\item{Focusing on applications to Gamma-ray bursts, in \cite{pw05}
a fully self-consistent, time-dependent homogeneous one-zone model 
was developed for the radiation from a 
relativistic plasma which assumes electron and proton injection into 
a power-law distribution and includes the self-consistent cooling of 
protons by $p\gamma$ pion production processes and their contributions 
to the pair populations (and their radiative output).}

\item{In the hadronic synchrotron mirror model \cite{boettcher05,reimer05}, 
developed specifically to explain the ``orphan'' TeV flare phenomenon
in 1ES~1959+650 mentioned above, the primary, correlated X-ray and TeV 
flare is explained by a standard SSC model while the secondary TeV-flare 
is explained by $\pi^0$-decay $\gamma$-rays as a result of photomeson 
production from relativistic protons interacting with synchrotron 
photons that have been reflected off clouds located at pc-scale 
distances from the central engine. Since some of the numerical values
in the original paper \cite{boettcher05} were in error (see the erratum),
this model will be revisited and re-evaluated in the next section.
}

\end{itemize}

\section{\label{sy_mirror}The hadronic synchrotron mirror model revisited}

\begin{figure}[t]
\includegraphics[width=0.48\textwidth]{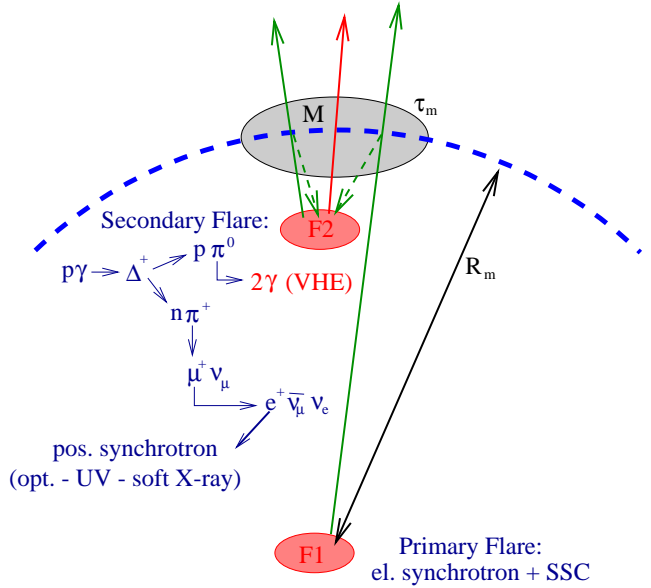}
\caption{Geometry of the model. A primary synchrotron flare is produced
by the emission region near the center of the system (F1). Synchrotron
emission is reflected at the mirror (M), and re-enters the emission
region. A quiescent level of synchrotron emission towards the mirror
will be sustained following the initial synchrotron flare. Its reflection 
into the emission region is the primary source of external photons 
leading to the secondary, ``orphan'' TeV flare as the emission region 
approaches the mirror (F2).}
\label{geometry}
\end{figure}

The recent ``orphan'' TeV flare of 1ES~1959+650 led to the development
of the hadronic synchrotron mirror model \cite{boettcher05,reimer05}.
The basic model geometry is sketched in Fig. \ref{geometry}. A blob 
filled
with ultrarelativistic electrons and relativistic protons is 
traveling along
the relativistic jet, defining the positive $z$ axis. 
Particles are primarily accelerated
very close to the central engine 
(F1) in an explosive event which is producing
the initial synchrotron 
+ TeV flare via the leptonic SSC mechanism. Acceleration of relativistic 
particles is expected to persist throughout the further propagation of 
the emission region along the jet through any of the processes (internal 
shocks, shear layer boundary acceleration, etc.) mentioned in 
\S \ref{models}, thus leading to a sustained level of (quiescent 
state) UV/X-ray synchrotron emission, as observed in 1ES~1959+650
\cite{krawczynski04}. A fraction of this synchrotron
radiation is reflected off a gas cloud (the mirror) located
at a distance 
$R_m$ from the central engine. For the sake of analytical
tractability, 
we assume that the mirror (M) is a homogeneous shell with a
reprocessing 
depth $\tau_m = 0.1 \, \tau_{-1}$. 

The observed time delay between the primary synchrotron flare and 
the
secondary flare due to interactions of the blob with the first 
reflected
synchrotron flare photons to re-enter the blob 
was $\Delta t_{\rm obs} = 20 \, \Delta t_{20}$~days,
and 
is related to the distance of the reflector by
$\Delta t_{\rm obs} 
\approx {R_m \over 2 \, \Gamma^2 c}$. Thus, $R_m \approx 3 \Gamma_1^2 
\, \Delta t_{20}$~pc. As shown in \cite{boettcher05}, a cloud of reflecting
gas of typical radial extent $\Delta r = 10^{17} \, \Delta r_{17}$~cm and 
density $n_c = 10^6
\, n_6$~cm$^{-3}$ at this distance from the central 
source will result in a negligible flux in emission lines, and the 
expected duration of the secondary flare is $w_{\rm fl}^{\rm obs} \sim
1.2 \, \Gamma_1^{-2}$~hr, consistent with the observed time profile of 
the secondary TeV flare in 1ES~1959+650.

From the observed $\nu F_{\nu}$ fluxes in synchrotron and TeV emission
during the secondary TeV flare, $\nu F_{\nu} ({\rm sy}) \sim 2 \times
10^{-10}$~ergs~s$^{-1}$~cm$^{-2}$ and $\nu F_{\nu} (600 \, {\rm GeV})
\sim 3 \times 10^{-10}$ ergs~s$^{-1}$~cm$^{-2}$ \cite{krawczynski04},
we find the co-moving luminosities, ${L'}_{\rm sy} \sim 1.0 \times
10^{41} \, \Gamma_1^{-4}$~ergs~s$^{-1}$ and ${L'}_{\rm VHE} \sim 1.5
\times 10^{41} \, \Gamma_1^{-4}$~ergs~s$^{-1}$. With a blob radius of
${R'_B} = 10^{16} \, R_{16}$ cm, this yields a co-moving synchrotron
radiation field of 

\begin{eqnarray}
& {u'}_{\rm sy} & \sim {9 \, d_L^2 \over 4 \, {R'}_B^2 c \, \Gamma^4} \, 
\nu F_{\nu} ({\rm sy})
\cr\cr
& & \sim 6.0 \times 10^{-3} \, \Gamma_1^{-4} \, 
R_{16}^{-2} \; {\rm ergs
\, cm}^{-3}.
\label{u_sy}
\end{eqnarray}
The characteristics of the reflected synchrotron flux (with co-moving
energy density ${u'}_{\rm Rsy}$) have been calculated in \cite{bd98}. 
Using their Eq. (4) in the limit $z \gg 2 \, \Gamma \, {R'}_B$, we find

\begin{equation}
{u'}_{\rm Rsy} \sim 0.24 \, \Gamma_1^{-1} \, \tau_{-1}
\, 
(\Delta r_{17}^{-1}) \; {\rm ergs \, cm}^{-3}.
\label{u_Rsy}
\end{equation}
The co-moving luminosity from $p\gamma \to \Delta \to p+\pi^0 \to p +
2 \gamma$ produced by protons of a given energy ${\gamma'}_p$ is then
given by

\begin{equation}
{L'}_{\rm VHE} \sim {8 \over 3} \, c \, \sigma_{\Delta} \, {u'}_{\rm Rsy}
{\gamma'}_p \, \Delta {\gamma'}_p \, {70 \, {\rm MeV} \over {E'}_{\rm Rsy}} 
\, {N'}_p ({\gamma'}_p).
\label{LRsy}
\end{equation}
where $\sigma_{\Delta} \approx 300 \, \mu$b is the $\Delta$ resonance peak
cross
section, $\Delta {\gamma'}_p \sim {\gamma'}_p/2$ parametrizes the
FWHM of the $\Delta$ resonance, and ${N'}_p ({\gamma'}_p)$ is the differential
number of protons at energy
${\gamma'}_p$. With this, the observable 
$\nu F_{\nu}$ peak flux in the
TeV flare can be estimated as

\begin{eqnarray}
\nu F_{\nu} ({\rm VHE}) & \sim & {{L'}_{\rm VHE} \, \Gamma^4 \over 4 \pi \,
d_L^2} \cr\cr
& \sim & 3.6 \times 10^{-59} \, {N'}_p ({\gamma'}_p) \,
E_{\rm sy, 1}^{-3} \cr\cr
& & \tau_{-1} \, R_{16}^{-1} \, (\Delta r_{17})^{-1}
\; {\rm ergs \, cm}^{-2}
\, {\rm s}^{-1}.
\label{nFn_VHE}
\end{eqnarray}
Setting this equal to the observed VHE peak flux, we find

\begin{equation}
{N'}_p ({\gamma'}_p) \sim 8.3 \times 10^{48} \, E_{\rm sy, 1}^{3} \, 
\tau_{-1}^{-1} \, R_{16} \, \Delta r_{17}.
\label{Np_gp}
\end{equation}
If the non-thermal proton spectrum is a straight power-law with energy
index $s = 2$ and low-energy cut-off ${\gamma'}_{\rm p, min} = \Gamma$,
this corresponds to a total proton number of ${N'}_{p, total} \sim
7.5 
\times 10^{55} \Gamma_1^{-3} \, \tau_{-1}^{-1} \, E_{\rm sy, 1} \, R_{16} 
\, \Delta r_{17}$
and a proton number density of 
\begin{equation}
{n'}_p \sim 1.8 \times 10^7 \, \Gamma_1^{-3} \,
E_{\rm sy, 1} \,
\tau_{-1}^{-1} \, R_{16}^{-2} \, \Delta r_{17} \; {\rm cm}^{-3}.
\label{np}
\end{equation}
We note that, in order to bring this to a value in the range of electron
densities typically invoked for leptonic jet models of blazar emission
(${n'}_e \sim$~a few $10^3$ -- $10^4$), a substantially higher Doppler
factor, a much more compact mirror, or a flatter proton spectrum,
$s < 2$, seems to be required. From Eq. \ref{np}, we find the total 
energy in
relativistic protons in the blob as ${E'}_{b,p} \, \sim \, 8.5 
\times
10^{47} \, \Gamma_1^{-2} \, E_{\rm sy, 1} \, \tau_{-1}^{-1}
\, R_{16} \, \Delta r_{17}$~erg.
If blobs of such relativistic plasma
fill a fraction $f = 10^{-3} \, f_{-3}$ of the jet, this corresponds
to a kinetic luminosity in protons of $L_p \sim
1.8 \times 10^{48} 
f_{-3} \, E_{\rm sy, 1} \, \tau_{-1}^{-1} \, \Delta r_{17}$~ergs~s$^{-1}$. 
These numbers indicate that orphan TeV flares as observed in 1ES~1959+650
and Mrk~410 seem to require some rather extreme conditions which might 
only be present at very rare occasions. This may explain why there have
so far only been very few examples observed.

\begin{acknowledgements}
This work has been supported by NASA through INTEGRAL GO (Theory)
grant NNG~05GK95G, INTEGRAL GO grant NNG~06GD57G, and by the
Harvard-Smithsonian Astrophysical Observatory through Chandra 
GO grant GO6-7101A. 
\end{acknowledgements}

\end{document}